\documentstyle[preprint,prd,aps]{revtex}
\begin{document}
\draft
%\preprint{ITP-96-20}
\pagenumbering{roma}
%%%%%%%%%%%%%%%%%%%%%%%%%%%%%%%%%%%%%%%%%%%%%%%%%%%%%%%%%%%%%%%%%%%%%%%%%%%%%%%%
\author{Yuan-Ben Dai, Chao-Shang Huang, Ming-Qiu Huang, Chun Liu}
\address{Institute of Theoretical Physics, Academia Sinica, P.O.Box 2735, Beijing 100080, China}
\title{QCD Sum Rules for Masses of Excited Heavy Mesons}
%\date{\today}
\maketitle
\thispagestyle{empty}
\vspace{15mm}
\begin{abstract}
The masses of excited heavy mesons are studied with sum rules in the heavy 
quark effective theory. A set of interpolating currents creating
(annihilating) excited heavy mesons with arbitrary spin and parity are
proposed and their properties are discussed. Numerical results at the
leading order of the ${\cal O}(1/m_Q)$ expansion are obtained for the lowest doublets $(0^+, 1^+)$ and $(1^+, 2^+)$.
\\

\noindent
%{\bf PACS} number(s): 14.20.Lq, 12.39.Hg, 13.30.Eg\\
\end{abstract}

\newpage
\pagenumbering{arabic}
%%%%%%%%%%%%%%%%%%%%%%%%%%%%%%%%%%%%%%%%%%%%%%%%%%%%%%%%%%%%%%%%%%%%%%%%%%%%%%%%
\pagenumbering{arabic}
Properties of excited heavy mesons composed of a heavy quark and a light anti-quark are of interest for several reasons. Some of these excited states have been observed in experiments. They will be objects of further study in 
future experiments with B-factories. In particular, they are useful for
 tagging in CP experiments \cite{eichten}. Theoretically, the relative simplicity of the dynamical condition makes it of interest for exploring the internal dynamics of systems containing a light quark.

The $1/m_Q$ expansions are useful for such systems ($m_Q$ is the heavy quark mass). The heavy quark effective theory (HQET) \cite{grinstein} is a powerful tool. However, for obtaining some detailed predictions one needs to combine it with some non-perturbative methods.
The spectra and decay widths of heavy meson excited states have been studied with the $1/m_Q$ expansion in the relativistic Bethe-Salpeter equations in [3,4]. They can also be studied with the QCD sum rules in HQET which have been used for ground states of heavy mesons[5,6].

As a first step, in the present article we shall use the QCD sum rules in HQET to obtain the masses of lowest heavy meson doublets $(0^+,1^+)$ and $(1^+,2^+)$. These states have been studied with the sum rules in the full QCD in the literature \cite{colangelo}. We shall see in the following
that using the HQET has some advantage for such problems.

In \cite{dai3} it was shown that if the interaction kernel contains only $\gamma_{\mu}$ and $I$ on the heavy quark line, solutions
of the Bethe-Salpeter equation for heavy mesons of arbitrary spin and definite parity
consist of two series of degenerate spin doublets in the infinity quark mass limit
$m_Q\to\infty$. The wave functions $\chi_{jPi}$, where $j$, $P$ are spin and parity
and $i=1, 2$ for these two series of states, have the following most general form

\begin{mathletters}
\label{wave1}
\begin{eqnarray}
\chi_{j,(-1)^{j+1},1}(x,u)&=&e^{im_Bv\cdot x}\sqrt{\frac{2j+1}{2j+2}}\:\frac{1+\not\!v}{2}\gamma^5\eta_{\alpha_1\cdots\alpha_j}(-i)^j\frac{\partial}{\partial u_{t\alpha_2}}
\cdots\frac{\partial}{\partial u_{t\alpha_j}}\nonumber\\
&&\times\:(\frac{\partial}{\partial u_{t\alpha_1}}-
{\frac{j}{2j+1}}\gamma_t^{\alpha_1}{\frac{\not\!\partial}{\partial u_{t}})
(\Phi_{1j}+i\frac{\not\!\partial}{\partial u_{t}}\Phi_{2j}})\;,\\[2mm]
\chi_{j+1,(-1)^{j+1},1}(x,u)&=&e^{im_Bv\cdot x}\frac{1}{\sqrt{2}}\frac{1+\not\!v}{2}\eta_{\alpha_1\cdots\alpha_{j+1}}\gamma_t^{\alpha_1}(-i)^j\frac{\partial}{\partial u_{t\alpha_2}}
\cdots\frac{\partial}{\partial u_{t\alpha_{j+1}}}\nonumber\\
&&\times\:(\Phi_{1j}+i\frac{\not\!\partial}{\partial u_{t}}{\Phi_{2j}})\;,
\end{eqnarray}
\end{mathletters}
and
\begin{mathletters}
\label{wave2}
\begin{eqnarray}
\chi_{j+1,(-1)^{j},2}(x,u)&=&e^{im_Bv\cdot x}\frac{1}{\sqrt{2}}\frac{1+\not\!v}{2}\gamma^5\eta_{\alpha_1\cdots\alpha_{j+1}}\gamma_t^{\alpha_1}(-i)^j\frac{\partial}{\partial u_{t\alpha_2}}
\cdots\frac{\partial}{\partial u_{t\alpha_{j+1}}}\nonumber\\
&&\times\:(\Psi_{1j}+i\frac{\not\!\partial}{\partial u_{t}}{\Psi_{2j}})\;,\\[2mm]
\chi_{j,(-1)^{j},2}(x,u)&=&e^{im_Bv\cdot x}\sqrt{\frac{2j+1}{2j+2}}\:\frac{1+\not\!v}{2}\eta_{\alpha_1\cdots\alpha_j}(-i)^j\frac{\partial}{\partial u_{t\alpha_2}}
\cdots\frac{\partial}{\partial u_{t\alpha_j}}\nonumber\\
&&\times\:(\frac{\partial}{\partial u_{t\alpha_1}}-
{\frac{j}{2j+1}}\gamma_t^{\alpha_1}{\frac{\not\!\partial}{\partial u_{t}})
(\Psi_{1j}+i\frac{\not\!{\partial}}{\partial u_{t}}\Psi_{2j}})\;.
\end{eqnarray}
\end{mathletters}
In the above formulas, $m_B$ is the meson mass, $x$ is the coordinate of the heavy quark,
$u$ is the coordinate of the light anti-quark relative to the heavy quark, $v$ is the velocity of the heavy meson, $u_l=v\cdot u$, $u_{t\mu}=u_{\mu}-u_lv_{\mu}$,
$\gamma_{t\mu}=\gamma_{\mu}-\not\! v v_{\mu}$, $\eta_{\alpha_1\cdots\alpha_{j}}$
is the transverse, symmetric and traceless polarization tensor. Finally, $\Phi_i$ and $\Psi_i$ are invariant functions of $u^2_t$ and $u_l$.

The assumptions used to obtain these results can be slightly generalized compared to that explicitly written down in \cite{dai3}. We only need to assume that in the $m_Q\to\infty$ limit the kernel is of the form

\begin{equation}
\label{kernel}
G(P,p,p')=I\otimes\Gamma \:U_S(P,p,p')+\gamma^{\mu}\otimes\Gamma_{\mu}\:U_V(P,p,p')
\end{equation}
where $P$ is the total momentum, $p$ and $p'$ are the relative momenta of the quarks in the initial and final state respectively, $\Gamma$ and $\Gamma_{\mu}$ are arbitrary functions of momenta and $\gamma$ matrices for the light quark which
transform as a scalar and a vector respectively, $U_S$ and $U_V$ are arbitrary
invariant functions of momenta.
Instead of the free propagator, the  general form can be used for quark propagators appearing in the B-S equation.
The calculations leading to above results only need to be slightly changed for
this general case. The restriction to $I$ and $\gamma_{\mu}$ in (\ref{kernel}) for the heavy
quark is equivalent to that the interaction is independent of the heavy quark
spin in the limit $m_Q\to\infty$. This is satisfied by QCD.\footnote{In QCD the elementary interaction vertex of the quark with the gluon 
is $\gamma_{\mu}$, the heavy quark propagator is $\displaystyle{{1\over v\cdot k}{1+\not\! v\over 2}}$ in the $m_Q\to\infty$ limit. Using $\displaystyle{{1+\not\! v\over 2}\gamma_{\mu}{1+\not\! v\over 2}=v_{\mu}{1+\not\! v\over 2}}$, the
 contribution of any Feynman diagram to the B-S kernel can be reduced to the form (\ref{kernel}) after integrating out the internal momentum of the diagram. } 
Besides this, the interaction kernel is most general. The B-S wave function is
not gauge invariant so that functions $\Phi_i$ and $\Psi_i$ are gauge dependent.
  However, the general form of (\ref{wave1}) and (\ref{wave2}) is the same
 for any gauge. Physically, each doublet has a definite value for the total
 spin of the light component. Except for the states $0^-$ and $0^+$, for each
 state in the series (\ref{wave1}) there is a state with the same j, P in the
 series (\ref{wave2}). They correspond to different values of $\displaystyle
 {j_l=j+\frac{1}{2}}$ and $\displaystyle{j-\frac{1}{2}}$. In the following we shall switch between the indices $i$ and $j_l$. This pattern is in agreement with the heavy quark symmetry. 
The above results are model independent consequences of the heavy
 quark symmetry. The B-S equation, being Lorentz covariant, only serves as
a convenient tool for finding the form of the wave function of the state
with definite values for the angular momenta $j$, $j_l$ and the 
parity $P$. For related results, see \cite{hussain,falk}. The authors
of these latter works used directly the transformation properties of the
wave function of a composite system with the angular momenta $j$ and $j_l\,$.
\cite{falk} contains results similar to ours, though the author did not
deal exactly with the B-S wave function. This again implies that our results
on the forms of the B-S wave functions in the $m_Q\to\infty$ limit are 
model independent.
 
 The relations between B-S wave functions and the matrix elements of the currents make the above results useful for constructing the appropriate interpolating currents and deriving their properties.
The wave function $\chi_{jPi}$ in (\ref{wave1}) and (\ref{wave2}) can be written in the form
\begin{equation}
\chi_{j,P,i}=e^{im_Bv\cdot x}\frac{1+\not\!v}{2}\eta_{\alpha_1\cdots\alpha_j}
\Gamma_{j,P,i}^{\alpha_1\cdots\alpha_j}(\frac{{\partial}}{\partial u_{t}})\left(f_1(u_t^2,u_l)+i\frac{\not\!{\partial}}{\partial u_{t}}f_2(u_t^2,u_l)\right)\;.
\end{equation}
We propose to use 
\begin{mathletters}
\label{current}
\begin{eqnarray}
\label{current1}
J_{j,P,i}^{\dag\alpha_1\cdots\alpha_j}(x)=\bar h_v(x)\Gamma_{j,P,i}^{\{\alpha_1\cdots\alpha_j\}}({\cal {D}}_{x_t})q(x)\;,
\end{eqnarray}
or
\begin{eqnarray}
\label{current2}
J_{j,P,i}^{'\dag\alpha_1\cdots\alpha_j}(x)=\bar h_v(x)\Gamma_{j,P,i}^{
\{\alpha_1\cdots\alpha_j\}}({\cal {D}}_{x_t})(-i)\not\!{\cal D}_{x_t}q(x)\;,
\end{eqnarray}
\end{mathletters}
as the interpolating current which creates the corresponding heavy meson, where $\displaystyle{h_v(x)=e^{im_Bv\cdot x}\frac{1+\not\!v}{2}Q(x)}$ is the heavy quark field in HQET and $Q(x)$ is that in full QCD, ${\cal {D}}_{x_t}$ is the covariant derivative $\displaystyle{\frac{{\partial}}{\partial x_t}-igA_t(x)}$ and
\begin{eqnarray*}
\Gamma^{\{\alpha_1\cdots\alpha_j\}}({\cal {D}}_{x_t})=\text{Symmetrize}\left\{\Gamma^{\alpha_1\cdots\alpha_j}({\cal {D}}_{x_t})-
\frac{1}{3}g_t^{\alpha_1\alpha_2}g_{\alpha_1^{'}\alpha_2^{'}}^t
\Gamma^{\alpha_1^{'}\alpha_2^{'}\alpha_3\cdots\alpha_j}({\cal {D}}_{x_t})\right\}\nonumber\;
\end{eqnarray*}
with $g_t^{\alpha_1 \alpha_2}=g^{\alpha_1\alpha_2}-v^{\alpha_1}v^{\alpha_2}$.\\
The currents hermitian conjugate to (\ref{current1}) and (\ref{current2})
\begin{mathletters}
\label{conju}
\begin{eqnarray}
\label{conju1}
J_{j,P,i}^{\alpha_1\cdots\alpha_j}(x)&=&\bar q(x)\bar{\Gamma}_{j,P,i}^
{\{\alpha_1\cdots\alpha_j\}}(\stackrel{\leftarrow}{\cal {D}}_{x_t})h_{v(x)}\;,\\[2mm]
\label{conju2}
J_{j,P,i}^{'\alpha_1\cdots\alpha_j}(x)&=&\bar q(x)(i\stackrel{\leftarrow}
{\not\!{\cal D}}_{x_t})\bar{\Gamma}_{j,P,i}^{\{\alpha_1\cdots\alpha_j\}}(\stackrel{\leftarrow}{\cal {D}}_{x_t})h_{v(x)}\;,
\end{eqnarray}
\end{mathletters}
where $\bar\Gamma=\gamma^0\Gamma^{\dag}\gamma^0$, annihilate the same meson.
The currents defined in (5) and (6) have nice properties. Let $|j,P,i\rangle$ be the heavy meson state with the quantum numbers $j$, $P$, $i$ in the $m_Q\to\infty$ limit. We have
\begin{eqnarray}
\label{decay}
\langle 0|J_{j,P,i}^{\alpha_1\cdots\alpha_j}(0)|j',P',i'\rangle=f_{Pj_l}\delta_{jj'}\delta_{PP'}\delta_{ii'}\eta^{\alpha_1\cdots\alpha_j}\;,
\end{eqnarray}
where $f_{jPi}=f_{Pj_l}$ has the same value for the two states in the same doublet. The formula
(\ref{decay}) can be verified as the following. Without loss of generality we can choose a gauge satisfying $A_{\mu}(x_0)=0$. In such a gauge, it follows from
 the definition of the B-S wave function $\chi_{j,P,i}(x,u)=\langle 0|T\left
(Q(x)\bar q(x+u)\right )|j,P,i\rangle$ that for a current in (\ref{conju1})
\begin{eqnarray}
\label{element}
\langle 0|J_{j,P,i}^{\alpha_1\cdots\alpha_j}(x_0)|j',P',i'\rangle=-\text{tr}\left [\chi_{j',P',i'}(0,u)\:\bar{\Gamma}_{j,P,i}^{\{\alpha_1\cdots\alpha_j\}}
(\stackrel{\leftarrow}{\partial_{u_t}})\right ]_{u=0}e^{i\bar\Lambda v\cdot x_0}\hspace{3cm}\nonumber\\[2mm]
=-\text{tr}\left\{\bar{\Gamma}_{j,P,i}^{\{\alpha_1\cdots\alpha_j\}}(\partial_{u_t})
\frac{1+\not\!v}{2}\:\Gamma_{j',P',i'}^{\beta_1\cdots\beta_{j'}}(\partial_{u_t})\left [f_1(u_t^2,u_l)+i\not\!{\partial}_{u_t}f_2(u_t^2,u_l)\right ]\right\}_{u=0}\eta_{\beta_1\cdots\beta_{j'}}e^{i\bar\Lambda v\cdot x_0}\;,
\end{eqnarray}
where $\bar\Lambda=m_h-m_Q$. Using the forms of $\Gamma$ in (1) and (2) and the relations of the type
\begin{mathletters}
\begin{eqnarray}
\label{ss}
\frac{\partial}{\partial u_{t\alpha_1}}
\cdots\frac{\partial}{\partial u_{t\alpha_{2l+1}}}f(u_t^2,u_l)|_{u_t=0}=0\;,\hspace{2.5cm}\\[2mm]
{\cal S}\left (\frac{\partial}{\partial u_{t\alpha_1}}
\cdots\frac{\partial}{\partial u_{t\alpha_{j}}}\right )\left (\frac{\partial}{\partial u_{t\beta_1}}
\cdots\frac{\partial}{\partial u_{t\beta_{j}}}\right )f(u_t^2,u_l)|_{u_t=0}\nonumber\\[2mm]
=\frac{\partial^j f}{(\partial u_t^2)^j}\:2^jj!\:{\cal S}\:g_t^{\alpha_1\beta_1}\cdots
g_t^{\alpha_j\beta_j}\delta_{jj'}|_{u_t=0}\;,
\end{eqnarray}
\end{mathletters}
where $\cal S$ denotes symmetrization and subtracting the trace terms in the sets $(\alpha_1\cdots\alpha_j)$ and $(\beta_1\cdots\beta_{j'})$, (\ref{decay}) can be
obtained from (\ref{element}). Only the term $f_1$ contributes to $f_{Pj_l}$.
If instead of $J$ we use $J'$ in (\ref{conju2}), the calculation is almost the same, except that now only the $f_2$ term contributes to $f_{Pj_l}$.

Using the equation of motion for the light quark $i {\not\!{\cal D}}q=0$ and
that for the heavy quark in HQET $iv\cdot Dh_v=0$, one can obtain the relation
\begin{equation}\label{r}
f^\prime_{jPi}=(-1)^{i+1} {\bar {\Lambda}}_{jPi}f_{jPi}
\end{equation}
between the coupling constants for the current with the extra factor
$i{\not\!{\cal D}}$ and that without it.

Similarly, calculated with the quark and gluon fields in the $m_Q\to\infty$ limit the correlation functions of $J_{j,P,i}$ and $J^{\dag}_{j',P',i'}$ satisfy
\begin{eqnarray}
\label{corr}
i\:\langle 0|T\left (J_{j,P,i}^{\alpha_1\cdots\alpha_j}(x)J_{j',P',i'}^{\dag\beta_1\cdots\beta_{j'}}(0)\right )|0\rangle&=&\delta_{jj'}\delta_{PP'}\delta_{ii'}(-1)^j\:{\cal S}\:g_t^{\alpha_1\beta_1}\cdots g_t^{\alpha_j\beta_j}\nonumber\\[2mm]&&\times\:\int \,dt\delta(x-vt)\:\Pi_{P,j_l}(x)\;,
\end{eqnarray}
where $\Pi_{P,j_l}$ is the same function for the two states in the same doublet.
For verifying (\ref{corr}) we use the Schwinger gauge
\begin{eqnarray}
\label{gauge}
x^{\mu}A_{\mu}=0.
\end{eqnarray}
The heavy quark propagator in HQET is 
\begin{eqnarray}
\label{propagator}
i\:\langle 0|T\left (h_v(x)\bar h_v(0)\right )|0\rangle=\frac{1+\not\!v}{2}\:\int\,
dt\delta^4(x-vt)\;.
\end{eqnarray}
The leading order Lagrangian in HQET is ${\cal L}_0=i\bar h_vv\cdot{\cal D}h_v$.
Therefore in gauge(\ref{gauge}) there is no gluon line terminating on the heavy quark line. We thus obtain
\begin{eqnarray}
\label{corr1}
i\:\langle 0|T\left (J_{j,P,i}(x)J_{j',P',i'}^{\dag}(0)\right )|0\rangle
&=&-i\int dt\delta^4(x-vt)\langle 0|T\left\{\text{tr}_{(color)}\text{tr}\left[\frac{1+\not\!v}{2}\:\right.\right.\nonumber\\[2mm]
&&\left.\left.\times\Gamma_{j',P',i'}({\cal D}_{yt})q(y)\bar q(x)\bar\Gamma_{j,P,i}(\stackrel{\leftarrow}{\cal D}_{x_t})\right]\right\}_{y=0,x=vt}.
\end{eqnarray}
Using the relation
\begin{eqnarray*}
\partial_{x_t\alpha_1}\cdots\partial_{x_t\alpha_j}\partial_{y_t\beta_1}\cdots\partial_{y_t\beta_{j'}}\text{tr}_{(color)}\left\{q(y)\bar q(x)exp\left[ig\int_y^xdz^{\mu}A_{\mu}\right]\right\}_{y=0,x=vt}\hspace{1.5cm}\\
=\text{tr}_{(color)}\left\{{\cal D}_{y_t\beta_1}\cdots{\cal
D}_{y_t\beta_{j'}}q(y)\bar q(x)\stackrel{\leftarrow}{{\cal
D}}_{xt\alpha_1}\cdots\stackrel{\leftarrow}{{\cal D}_{xt\alpha_j}}\right\}_{y=0,x=vt}\;,
\end{eqnarray*}
in the gauge (\ref{gauge}), one can rewrite (\ref{corr1}) in the form
\begin{eqnarray}
\label{corr2}
i\:\langle 0|T\left (J_{j,P,i}(x)J_{j',P',i'}^{\dag}(0)\right )|0\rangle
&=&-i\int dt\delta^4(x-vt)\text{tr}\left\{\Gamma_{j',P',i'}\right.\nonumber\\[2mm]&&\left.\times(\partial_{y_t}){\large S}(y,x)\bar\Gamma_{j,P,i}(\stackrel{\leftarrow}{\partial}_{x_t})\frac{1+\not\!v}{2}\right\}_{y=0,x=vt}\;,
\end{eqnarray}
where 
\begin{eqnarray}
{\large S}(y,x)=\text{tr}_{(color)}\langle 0|T\left (q(y)\bar q(x)exp\left[ig\int_y^xdz^{\mu}A_{\mu}\right]\right)|0\rangle\;.
\end{eqnarray}
From Lorentz and translation invariance for gauge invariant quantities, 
${\large S}(y,x)$ has the general form
\begin{eqnarray}
\label{s1}
{\large S}(y,x)\frac{1+\not\!v}{2}=\left[{\large S}_1((y-x)^2_t,y_l-x_l)+i\not\!\partial_{x_t}{\large S}_2((y-x)^2_t,y_l-x_l)\right]\frac{1+\not\!v}{2}\;.
\end{eqnarray}
Substituting (\ref{s1}) into (\ref{corr2}) brings it into a form almost identical to
the $r.h.s.$ of (\ref{element}). The formula (\ref{corr})  then follows after similar calculations.

(\ref{decay}) and (\ref{corr}) imply that two currents in the sets (\ref{current1}) and (\ref{current2}) with not identical values of $j$, $P$, $i$ never mix in the $m_Q\to\infty$ limit either in the heavy meson or in the quark-gluon level and heavy quark symmetry is explicit with these currents. This verifies that they are the appropriate interpolating currents for heavy meson states with definite $j$, $P$ and the
light quark momentum $j_l$ which are conserved in QCD in $m_Q\to\infty$ limit.

The properties for currents mentioned above are important for applications to QCD sum rules for excited heavy mesons. If one use other currents or study the sum rules in full QCD, there are in general contributions from two nearby poles corresponding to states of the
same $j$, $P$. Their contributions may not be separated correctly. Furthermore, the mixing
of such two states can be calculated within our formalism by introducing the
 ${\cal O}(1/m_Q)$ terms in the Lagrangian in HQET.

For obtaining QCD sum rules in the HQET we study the correlator
\begin{eqnarray}
\label{sum1}
\Pi^{\alpha_1\cdots\alpha_j,\beta_1\cdots\beta_j}_{j,P,i}(k)=i\int d^4xe^{ik\cdot x}\langle 0|T\left(J_{j,P,i}^{\alpha_1\cdots\alpha_j}(x)J_{j,P,i}^{\dag\beta_1\cdots\beta_{j}}(0)\right )|0\rangle\;.
\end{eqnarray}
From (\ref{element}), it has the following form at the leading order
\begin{eqnarray}
\label{sum2}
\Pi^{\alpha_1\cdots\alpha_j,\beta_1\cdots\beta_j}_{j,P,i}(k)=(-1)^j\;{\cal S}\;
g_t^{\alpha_1\beta_1}\cdots g_t^{\alpha_j\beta_j}\:\Pi_{j,P,i}(\omega)\;,
\end{eqnarray}
where $\omega=2v\cdot k$ and $\Pi_{j,P,i}(\omega)=\Pi_{P,j_l}(\omega)$ is independent of $j$.

We shall confine us to the leading order terms in this article. The contribution to $\Pi_{P,j_l}(\omega)$ of the pole term on the hadron side is
\begin{eqnarray*}
\frac{f^2_{P,j_l}}{2\bar\Lambda-\omega}\;.
\end{eqnarray*}
On the quark-gluon side of the sum rule, $\Pi_{P,j_l}(\omega)$ is calculated with the leading order Lagrangian ${\cal L}_0$ in the HQET. The Feynman diagrams contributing to (\ref{sum1}) are
 shown in Fig 1. For the sake of simplicity we consider the doublets $(0^+,1^+)$
 and $(1^+,2^+)$ in the following.

According to (\ref{current1}) and (\ref{current2}), there are two possible choices for currents for the doublet $(0^+,1^+)$, either
\begin{eqnarray}
\label{curr1}
J^{\dag}_{0,+,2}&=&\frac{1}{\sqrt{2}}\:\bar h_vq\;,\\
\label{curr2}
J^{\dag\alpha}_{1,+,2}&=&\frac{1}{\sqrt{2}}\:\bar h_v\gamma^5\gamma^{\alpha}_tq\;,
\end{eqnarray}
or
\begin{eqnarray}
\label{curr3}
J^{'\dag}_{0,+,2}&=&\frac{1}{\sqrt{2}}\:\bar h_v(-i)\not\!{\cal D}_tq\;,\\
\label{curr4}
J^{'\dag}_{1,+,2}&=&\frac{1}{\sqrt{2}}\:\bar h_v\gamma^5\gamma^{\alpha}_t(-i)\not\!{\cal D}_tq\;.
\end{eqnarray}
Similarly, there are two possible choices for the currents for the doublet $(1^+,2^+)$. One is
\begin{eqnarray}
\label{curr5}
J^{\dag\alpha}_{1,+,1}&=&\sqrt{\frac{3}{4}}\:\bar h_v\gamma^5(-i)\left(
{\cal D}_t^{\alpha}-\frac{1}{3}\gamma_t^{\alpha}\not\!{\cal D}_t\right)q\;,\\
\label{curr6}
J^{\dag\alpha_1,\alpha_2}_{2,+,1}&=&\sqrt{\frac{1}{2}}\:\bar h_v
\frac{(-i)}{2}\left(\gamma_t^{\alpha_1}{\cal D}_t^{\alpha_2}+\gamma_t^{\alpha_2}{\cal D}_t^{\alpha_1}-2g_t^{\alpha_1\alpha_2}\not\!{\cal D}_t\right)q\;.
\end{eqnarray}
Another choice is obtained by adding a factor $-i\not\!{\cal D}_t$ to (\ref{curr5}) and (\ref{curr6}). Note
that, without the last term in the bracket in (\ref{curr5}) the current
would couple also to the $1^+$ state in the doublet $(0^+,1^+)$ even in
the limit of infinite $m_Q$.

Usually, the currents with the least number of derivatives are used in
the QCD sum rule approach. The sum rules with them have better convergence in the high
energy region and often have better stability. However, there is a motivation for using the
currents (\ref{curr3}), (\ref{curr4}) for the doublet $(0^+,1^+)$. In the
non-relativistic quark model, which usually gives correct ordering of 
energy levels of hadron states, the doublets $(0^+,1^+)$ and $(1^+,2^+)$
are orbital p-wave states which correspond to one derivative in the space
wave functions. In fact, as mentioned above, the coupling constant for the
currents (\ref{curr3}), (\ref{curr4}) is proportional to $\Psi_2$ in (2)
and that for the currents (\ref{curr5}), (\ref{curr6}) is proportional to
$\Phi_1$ in (1) which are the large components in the non-relativistic
approximation. Therefore, we shall consider both the currents
(\ref{curr1}), (\ref{curr2}) and (\ref{curr3}), (\ref{curr4}) for the
doublet $(0^+,1^+)$.

First we consider the doublet $(0^+,1^+)$. For the correlator
\begin{eqnarray}
\label{pi2}
\Pi_{0,+,2}=i\int d^4xe^{ik\cdot x}\langle 0|T\left({J'}_{0,+,2}(x){J'}_{0,+,2}^{\dag}(0)\right )|0\rangle\;,
\end{eqnarray}
where ${J'}_{0,+,2}$ is the current conjugate to (\ref{curr3}), the sum rules after the Borel transformation is found to be
\begin{eqnarray}
\label{form1}
&&f^2e^{-2\bar\Lambda/{T}}=\frac{3}{2^6\pi^2}\int_0^{\omega_c}\omega^4e^{-\omega/{T}}d\omega-\frac{1}{2^4}m_0^2\langle\bar qq\rangle\;,\\[3mm]
\label{form12}
&&\bar\Lambda=\frac{\displaystyle{\frac{3}{2^3\pi^2}\int_0^{\omega_c}\omega^5e^{-\omega/{T}}d\omega}}{\displaystyle{\frac{3}{4\pi^2}\int_0^{\omega_c}\omega^4e^{-\omega/{T}}d\omega-m_0^2\langle\bar qq\rangle}}\;.
\end{eqnarray}
In the above sum rule we have confined us to terms of lowest order in perturbation and operators of dimension less than six.

The correlator of the $1^+$ current ${J'}_{1,+,2}$ (\ref{curr4}) and
its conjugate has the form
\begin{eqnarray}
\label{pi1}
\Pi_{1,+,2}^{\alpha\beta}=i\int d^4xe^{ik\cdot x}\langle 0|T\left({J'}^{\alpha}_{1,+,2}(x){J'}^{\dag\beta}_{1,+,2}(0)\right )|0\rangle
=-g_t^{\alpha\beta}\:\Pi_{1,+,2}(\omega)\;.
\end{eqnarray}
 We find for $
\Pi_{1,+,2}(\omega)$ the identical sum rule as that for $\Pi_{0,+,2}
(\omega)$, as it should be.

The corresponding formula when the current $J_{0,+,2}$ and $J_{1,+,2}$ in (\ref{curr1}) and (\ref{curr2}) are used instead of
${J'}_{0,+,2}$ and ${J'}_{1,+,2}$ are the following 
\begin{eqnarray}
\label{form2}
&&f^2e^{-2\bar\Lambda/{T}}=\frac{3}{16\pi^2}\int_0^{\omega_c}\omega^2e^{-\omega/{T}}d\omega+\frac{1}{2}\langle\bar qq\rangle-{1\over 8T^2}m_0^2\langle\bar qq\rangle\;,\\[3mm]
\label{form22}
&&\bar\Lambda={\displaystyle{{3\over 2^4\pi^2}\int_0^{\omega_c}\omega^3e^{-\omega/{T}}d\omega
+{m_0^2\over 4T}\langle\bar qq\rangle}\over\displaystyle{{3\over 2^3\pi^2}\int_0^{\omega_c}\omega^2e^{-\omega/{T}}d\omega+\langle\bar qq\rangle-{1\over 4T^2}m_0^2\langle\bar qq\rangle}}\;
\end{eqnarray}
The correlators for the $(1^+,2^+)$ doublet have the form
\begin{eqnarray}
\label{pi3}
&&\Pi_{1,+,1}^{\alpha,\beta}(k)=-g_t^{\alpha\beta}\:\Pi_{+,j_l=3/2}(\omega)\;,\\[2mm]\label{pi4}
&&\Pi_{2,+,1}^{\alpha_1\alpha_2,\beta_1\beta_2}(k)={1\over 2}\left(g_t^{\alpha_1\beta_1}g_t^{\alpha_2\beta_2}+g_t^{\alpha_1\beta_2}g_t^{\alpha_2\beta_1}-{2\over 3}g_t^{\alpha_1\alpha_2}g_t^{\beta_1\beta_2}\right)\Pi_{+,j_l=3/2}(\omega)\;.
\end{eqnarray}
When the currents (\ref{curr5}) and (\ref{curr6}) are used the sum rules for
 this doublet are found to be
\begin{eqnarray}
\label{form3}
&&f^2e^{-2\bar\Lambda/{T}}={1\over 2^6\pi^2}\int_0^{\omega_c}\omega^4e^{-\omega/{T}}d\omega
-\frac{1}{12}\:m_0^2\:\langle\bar qq\rangle-{1\over 2^5}\langle{\alpha_s\over\pi}G^2\rangle T\;,\\
\label{form32}
&&\bar\Lambda=\frac{\displaystyle{\frac{1}{2^5\pi^2}\int_0^{\omega_c}\omega^5
e^{-\omega/{T}}d\omega-{1\over 2^4}\langle{\alpha_s\over\pi}G^2\rangle T^2}}
{\displaystyle{\frac{1}{2^4\pi^2}\int_0^{\omega_c}\omega^4e^{-\omega/{T}}d\omega-
{m_0^2\over 3}\langle\bar qq\rangle-{1\over 2^3}\langle{\alpha_s\over\pi}G^2\rangle T}}\;.
\end{eqnarray}
Here the $\langle{\displaystyle{\alpha_s\over\pi}}G^2\rangle$ term comes from the diagram (c) in Fig. 1
which vanishes for the doublet $(0^+,1^+)$.

From (\ref{form12}), (\ref{form22}) and (\ref{form32}) we obtain curves shown in
Fig. 2a,b,c. For the first two cases stability windows exist when
$\omega_c$ lies in the interval $2.4$ to $2.9$ GeV; for the last one $\omega_c$ lies in the interval $2.7$ to $3.2$ GeV. Imposing usual criterium
for the upper and lower bounds of the Borel transform parameter $T$ we found
approximately the same stability window $0.7$--$0.9$ GeV. The results for 
$\bar\Lambda$ for three cases are the following
\begin{equation}
\label{result1}
\bar\Lambda({1/2}^+)=1.05\pm 0.05 ~~\text{GeV},
\end{equation}
for the doublet $(0^+,1^+)$ when the currents
(\ref{curr1}), (\ref{curr2}) without the derivative is used.
\begin{equation}
\label{result2}
\bar\Lambda({1/2}^+)=0.90\pm 0.10 ~~\text{GeV},
\end{equation}
for the same doublet when the currents (\ref{curr3}), (\ref{curr4}) with
the derivative is used.
\begin{equation}
\label{result3}
\bar{\Lambda}({3/2}^+)=0.95\pm 0.10 ~~\text{GeV},
\end{equation}
for the doublet $(1^+,2^+)$.

The predicted mass value for the doublet $(0^+,1^+)$ in (\ref{result1}) is somewhat different from that in (\ref{result2}). The value in (\ref{result1}) is in agreement
with the result of [7] in which the same currents as (\ref{curr1}) and (\ref{curr2}) are used. (\ref{result1}) predicts a mass value of the doublet
$(0^+,1^+)$ slightly higher than the mass value of the doublet $(1^+,2^+)$ predicted
in (\ref{result3}). On the other hand, (\ref{result2}) gives a mass value for
the doublet $(0^+,1^+)$ approximately equal or slightly lower than that for the doublet $(1^+,2^+)$
in (\ref{result3}). 
Though (36) and (37) can be considered as consistent within the uncertainties, (37) indicates that the QCD sum rule approach can incorporate a mass value for $(0^+,1^+)$ lower than that for $(1^+,2^+)$. This is in agreement with quark model predictions in
[11] and [3]. By contrast, we have also derived the sum rule
for the ground state doublet $(0^-,1^-)$ with the currents with the
extra derivative $-i{\not\!{\cal D}}$. We found that there is no acceptable stability
window for this sum rule.

Finally, we would like to address the following points of this paper. The results from our previous work on B-S wave functions are used only for
finding the appropriate currents. Once these currents are given, the 
calculations in QCD sum rules are independent of the consideration of the B-S equation. That the currents used are appropriate was checked by proving (11)
with direct calculations in terms of quark-gluon fields in HQET without
referring to the B-S wave functions. 
Assuming duality and inserting intermediate heavy meson states between $J$ and $J^{\dagger}$ in (11), this equation also strongly suggests equation (7)
which is the direct statement that the currents create pure states with definite
values of $j$, $P$ and $j_l$. 

\acknowledgments

This work was completed when one of the
authors (Y.D.) was visiting the University Of Dortmund. Y.D. would like to thank Prof. Paschos and colleagues in his group for hospitality and helpful discussions. This work is supported in part by the National Natural Science Foundation Of
 China. C. L. is supported by the Postdoctoral Science Foundation of China.\\

\newpage
{\bf Figure Captions}
\vspace{2ex}
\begin{center}
\begin{minipage}{120mm}
{\sf
Fig. 1.} \small{Feynman diagrams contributing to the sum rule. There is no Fig (c) when the currents (20) (21) are used.}
\end{minipage}
\end{center}
\begin{center}
\begin{minipage}{120mm}
{\sf
Fig. 2.} \small{Dependence of $\bar\Lambda$ on the Borel parameter $T$ for different
values of the continuum threshold $\omega_c$. ({\large a})({\large b}) for $(0^+, 1^+)$ doublet with the interpolating currents
(\ref{curr3}) (\ref{curr4}) and (\ref{curr1}) (\ref{curr2}) used respectively. From top to bottom the curves correspond 
to ({\large a}) $\omega_c=2.8, 2.6, 2.4$ GeV and ({\large b}) $\omega_c=2.9, 2.65, 2.4$ GeV.
({\large c}) for doublet $(1^+, 2^+)$. From top to bottom  the curves correspond 
to $\omega_c=3.1, 2.9, 2.7$ GeV respectively. }
\end{minipage}
\end{center}
 

\vspace{1cm}
\begin{thebibliography}{1}

\bibitem{eichten}  E. Eichten, C. T. Hill and C. Quigg, Fermilab-Conf-94/118-T.

\bibitem{grinstein}  B. Grinstein, Nucl. Phys. {\bf B339}, 253(1990);
		    E. Eichten and B. Hill, Phys. Lett.  {\bf B234}, 511(1990);
		    A. F. Falk, H. Georgi, B. Grinstein and M. B. Wise, Nucl. Phys. {\bf B343}, 1(1990); F. Hussain, J. G. K\"{o}rner, K. Schilcher, G. Thompson and Y. L. Wu, Phys. Lett.  {\bf B249}, 295(1990); J. G. K\"{o}rner
and G. Thompson, Phys. Lett.  {\bf B264}, 185(1991).

\bibitem{dai1}  Y. B. Dai, C. S. Huang and H. Y. Jin, Phys. Lett.  {\bf B331}, 174(1994).          

\bibitem{dai2} Y. B. Dai and H. Y. Jin, Phys. Rev. {\bf D 52}, 236(1995). 

\bibitem{bagen}  E. Bagen, P. Ball, V. M. Braun and H. G. Dosch, Phys. Lett.  {\bf B278}, 457(1992); M. Neubert, Phys. Rev. {\bf D 45}, 2451(1992); D. J. Broadhurst and A. G. Grozin, Phys. Lett.  {\bf B274}, 421(1992); M. Neubert, Phys. Rep. {\bf 245}, 259(1994).

\bibitem{ball}  P. Ball and V. M. Braun, Phys. Rev. {\bf D 49}, 2472(1994).

\bibitem{colangelo}  P. Colangelo, G. Nardulli, A. A. Ovchinnikov and N. Paver, Phys. Lett.  {\bf B269}, 204(1991); P. Colangelo, G. Nardulli and N. Paver, Phys. Lett.  {\bf B293}, 207(1992).

\bibitem{dai3}  Y. B. Dai, C. S. Huang and H. Y. Jin, Zeit. Phys. {\bf C60}, 527(1993).

\bibitem{hussain} F. Hussain, G. Thompson, and J.G. K\"{o}rner,
	     ICTP report IC-93-314 (1993), Nov. 1993.

\bibitem{falk} A. F. Falk, Nucl. Phys. {\bf B378}, 79(1992).

\bibitem{isgur} S. Godfrey and N. Isgur, Phys. Rev. {\bf D 32}, 189(1985)

\end{thebibliography}
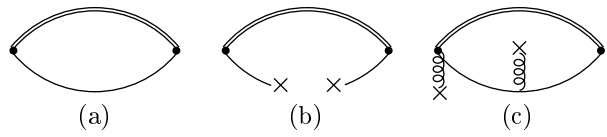
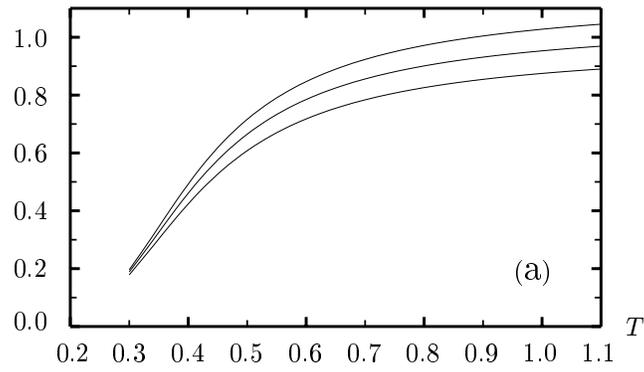
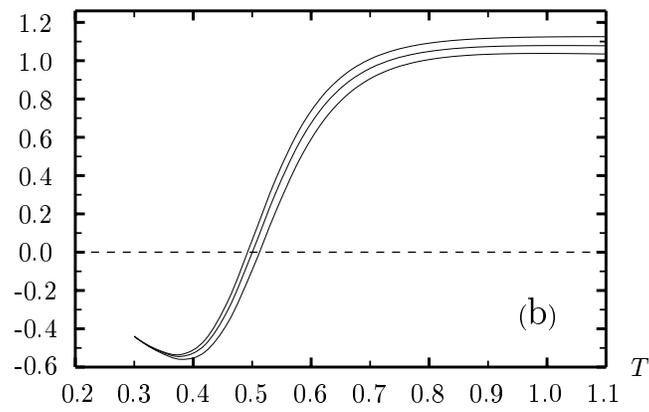
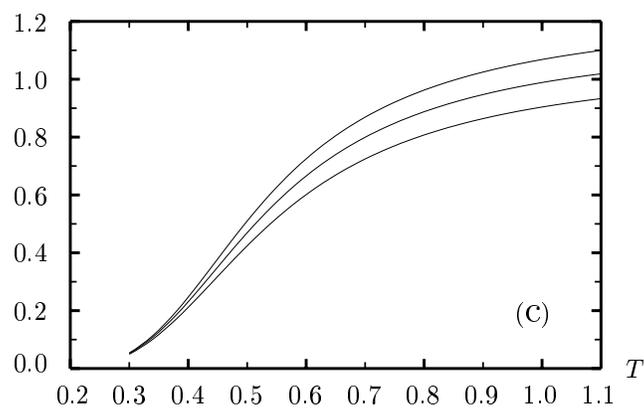
\end{document}